\documentclass{aa}
\usepackage{graphicx}
\begin{document}

%

%
   \title{Microflaring  of a solar  Bright Point 
\thanks{Based on observations with the {\it SUMER}-spectrometer on board
 the {\it SOHO} observatory}  }

  \author{O. Vilhu
\thanks{{\it SUMER} associated scientist }
         \inst{1}
       \and
        J. Huovelin \inst{1}
        \and
          S. Pohjolainen \inst{2}
         \and
          J. Virtanen \inst{1}
          \and
          W. Curdt 
\thanks{\it SUMER Co-I}
\inst{3}
          }

   \offprints{O. Vilhu}

   \institute{
Observatory, Box 14, FIN-00014 University of Helsinki, Finland\\
      \email{osmi.vilhu@helsinki.fi and huovelin@astro.helsinki.fi} 
\and
   Tuorla Observatory, University of Turku, Finland\\
          \email{spo@astro.utu.fi}
\and
      Max-Planck-Institut f\"ur Aeronomie, D-37191
       Katlenburg-Lindau, Germany\\
           \email{curdt@mpae.gwdg.de}
                                           }
   \date{Received  ; accepted  }


\abstract{ A 50 x 50 arc sec region near the solar disc center containing a 
bright point (BP)  
was observed with the {\it SUMER}-
spectrograph of the {\it SOHO} observatory. The data
consist of two hours observation of four far-UV emission lines formed 
between 2 10$^4$ --
6 10$^5$ K, with 2 arc sec spatial, 
2.8 min temporal  and 4 km/s  spectral
resolution. A striking feature was the strong microflaring of the
major persistent BP (with size 8 x 8 arcsec)  
 and the appearance of several short lived transients.  
The microflaring  of  each
individual 2 x 2 arc sec  pixel inside the main BP was coherent, 
indicating strong interaction
of the possible sub arc sec building blocks (magnetic flux tubes). 
Using  the emission measure at 10$^5$ K  as an indicator of the  
loop foot point area and magnetic filling 
factor, we
 suggest 10 per cent filling factor for the BP observed. This is similar to
that on the average  surface of a medium-active solar type  star.
    }
        \maketitle

\section{Introduction}

The nature of solar coronal bright points (BP) 
has been an enigma since their discovery in late 1960's
(e.g. Vaiana et al., 1970). Their correspondence with small
bipolar magnetic regions was found 
by combining ground-based magnetic field measurements with
simultaneous space-borne X-ray imaging observations
(Krieger et al., 1971; Golub et al., 1977). The bright points are clearly seen 
in Ca K spectroheliograms as bright ``knots'' (e.g. Golub \&
Pasachoff, 1997). The daily number of BP's found on the Sun 
varies between several hundreds up to a few thousand (Golub et al., 1974).
Zhang et al. (2001) found their density around 800 BP's for
the entire solar surface at any moment.

 Golub et al. (1974)
found that the diameters of the bright points in X-rays 
are around $10^{4}$~km (10 - 20 arc sec) and  
their lifetimes  range from 2 hours
to 2 days (see also Zhang et al. 2001). 
The physical studies have indicated the
temperatures to be fairly low, $T \approx 2 \times 10^6
$~K, and the electron densities  $n_e \approx 5 \times 10^9$~cm$^{-3}$
(e.g. Golub \& Pasachoff, 1997), although  cooler BP's exist
(Habbal, 1990). 
Assuming that almost
all BP's represent new magnetic flux emerging at the solar surface, 
their overall contribution to the solar 
magnetic flux would exceed that of the active regions 
(Golub and Pasachoff, 1997). 

BP's often show irregular intensity and spatial 
shape variations on time scales of 
minutes which can be called as microflares (Shimojo \& Shibata, 1999;
Zhang et al. 2001). In particular, Shimojo and Shibata show that the
frequency distribution of these microflares, as a function of the peak
intensity, show a power law with index 1.7$\pm{0.4}$ which is consistent
with that of ordinary (stronger) flares. However, the total  coronal heating
can not energetically be due to these microflares, they  are too few.
Brown et al. (2001), using {\it TRACE}, 
studied for the first time BP's for their
entire lifetime with a cadence of 2 min and spatial resolution of 0.5 arcsec,
using hot FeXII and FeXI lines. In particular, they suggested that BP's
are made up of a complex system of dense loops.

In the present paper we report observations of one particular bright point
 observed with the 
{\it SUMER}-spectrometer onboard {\it SOHO}  using four
ultraviolet lines formed between 2 10$^4$ - 6 10$^5$ K.  

\section{ Observations}

The observations were performed on October 4 and 8, 1996, with 
the SUMER normal incidence telescope of the {\it SOHO} observatory
(Wilhelm et al., 1995).
The full set of observations will be reported elsewhere (Huovelin et al. 2002);
here we concentrate on a specific BP-complex observed October 8, 1996, between
UT 15.20 - 17:20. 

The region was found by  first scanning  a 
larger area near the disk center.
The data consist of
$23 \times 24$~  raster scans of the detector B first
order. The scans were obtained by moving the spectrometer
slit horizontally
along the East-West (equatorial) direction in 2.25 arcsec steps,
corresponding to 1700~km on the solar surface.
The wavelength dispersion in each scan-point was perpendicular to the
South-North direction, and 4 narrow spectral windows 
were observed simultaneously in each vertical spatial point 
(with 2 arcsec height). The
observed wavelength intervals consist 
of 50 wavelength bins with a dispersion 
 0.0446-0.0447 \AA/px (17 km/s, comparable to the thermal broadening
of NIV 765 line formed at 10$^5$ K). The intervals were
centered at the lines N IV (765.146 \AA), Ne VIII (770.420 \AA), N II
(775.965 \AA) and O IV (787.740 \AA) (see Table 1). The resolving power
of {\it SUMER} within the observed wavelength range is close to $\lambda / \Delta
\lambda = 37000$ (8 km/s). For radial velocity measurements a better accuracy
(around 4 km/s) can be achieved.   
The scan time for each image was set to 2.8 minutes
(6.5~s integration per slit position plus overheads), resulting in 44 images for each spectral
line with spectral information inside each 2 x 2.25 arcsec spatial pixel.

The observations were reduced  
including flat-field  and geometrical distortion
corrections to the raw uncompressed FITS format data.  
These  standard reductions were made with the SUMER reduction
procedures written in IDL 
(Interactive Data Language, Registered Trademark)
at the Institute of Theoretical Astrophysics, University of Oslo,
Norway (M. Carlsson, private comm.).
Dead time correction was not applied, since the total signal level 
in the slit was below 50~000 cts/s (see the {\it SUMER} Data Cookbook,
http://www.mpae.gwdg.de/ mpae\_projects/ SUMER/text/cookbook.html).

The local gain correction was made to all spectra to compensate 
for gain variations.
Finally, wavelength scaling and 
radiometric conversion to physical units were applied.    
For the two shortest wavelengths (765 and 770
\AA) the scaling was  44.7 m\AA/pixel, and for the
other two, 44.6m\AA/pixel. 
The radiometric calibration was based on the results reported
by Wilhelm et al. (1997),  the latest calibrations available in the 
WWW-pages of {\it SUMER} during summer 1998 were used. The uncertainty
in the absolute flux calibration is $\pm 15$ \% (Sch\"uhle et al., 1998).
 The instrument compensated 
solar rotation during each sequence, and the spatial variability
in each region could be monitored.

\begin{table*}
\caption{\label{tab:line}\protect\small Lines observed. The formation
temperatures assume ionisation equilibria from Arnaud \& Rothenflug (1985).
The line intensities, widths (FWHM) and non-thermal velocities ($\xi$) represent measured mean values over
the 8 x 8 arcsec area and over the 120 min observation around the bright point A in Fig. 2.}            

\small
\begin{center}
\begin{tabular}{ccccccccc}
\multicolumn{9}{c}{ } \\ \hline

$\lambda$ (A)  &  Ion  & lower level & upper level   & temperature (K) & int (mW/Sr/m$^2$) & FWHM (km/s) & $\xi$ (km/s)  &                   \\
\hline

765.15 & NIV & 2s$^2$ $^1S_0$    &  2s2p $^1P_1$ & $10^5$                         &  305 $\pm{30}$  &  91 $\pm{5}$ &  44 $\pm{5}$ &    \\

770.42     &  NeVIII & $1s^2$2s $^2S_{1/2}$    & 1s$^2$2p $^2P_{3/2}$  & 6 $10^5$ & 200 $\pm{20}$   &  91 $\pm{5}$  &   41 $\pm{5}$ &                           \\

775.96     &  NII &  $2s^2$2p$^2$ $^1D_2$ &   2s2p$^3$ $^1D_2$ &   2 $10^4$      &  22 $\pm{5}$  &  82 $\pm{10}$  &    40 $\pm{10}$  &           \\

787.74   & OIV &   $2s^2$2p $^2P_{1/2}$ &  $2s2p^2$ $^2D_{3/2}$  &    2 $10^5$    &  240 $\pm{30}$  & 95 $\pm{5}$   &    46 $\pm{5}$  &       \\

\hline
\end{tabular}
\end{center}
\end{table*}

\section{Results} 

Line profiles  were
fitted with the IDL gaussfit-procedure, the mean values of resulting line 
parameters are given in Table 1. 
  These gaussian fits resulted in line intensity, line center 
 and line width (FWHM)
values at each spatial (23 x 24) and temporal (44, 
separated by 2.8 min) points. Although the spectral resolution was 8 km/s and
wavelength bins 17 km/s, the fitting procedure of strong lines 
permits to measure
the line centroids much better. Radial velocities and widths of
the strongest lines can be determined with accuracy better than 5 km/s
 (see Fig. 6).  
Average non-thermal velocities are also included in Table 1. They are slightly
below the sonic velocities and are situated close to  the upper
 boundary of 
the velocity -- temperature correlation in quiet sun 
as derived  by Dere \& Mason (1993, their Fig. 17).

 Fig.1 shows the  spectral windows observed. 
The quiet sun and spot spectra, as observed
with the same instrument settings, can be found from the {\it SUMER}-atlas
available via http://www.mpae.gwdg.de. Fig.2 shows the time-averaged
contour-plot of six
observed frames (frames numbers 29 - 34 in Fig.3).
The observed region was extremely variable as shown in Fig. 3 where every 
time-frame is presented. Fig. 4 shows  the light curve of the NIV 765
line for the
brightest and most persistent bright point 
in the observed region (A in Fig.2) indicating strong mictoflaring. 
The variability
  is very coherent with
strong correlation between individual 2 x 2 arc sec spatial pixels.
  If this bright point consisted of a complex
loop system (not resolved with our spatial resolution 
but suggested by {\it TRACE} observations of Brown et al. 2001) this would mean
a  rapid avalance over the whole bright point region after  triggering.

 The  line intensities were compared with those from
the {\it SUMER} ATLAS (quiet sun and spot spectra) by  dividing 
them  with those of the quiet sun. The line ratios reflect relative
emission measures and 
 are proportional to n$_e$$^2$ $\times$ V where
n$_e$ is the electron density and V the emitting volume at the line
formation temperature  
(see Fig. 5). The background in the 50 x 50 arcsec region 
 was somewhat different from the quiet
network e.g. by having an excess  hot component.

The observed region contained also some short lived transients, like 
BP-B  close to the main  BP-A (see Fig. 2). It showed clear
20 km/s up-flows  during the outburst (see Fig. 6). On the
other hand, BP-A did not show any variability in line shapes
 irrespective of it strong intensity variations.

\section{Discussion and Conclusions}

The main bright point (A in Fig. 2)  had dimensions
8 x 8 arcsec between 2 10$^4$ - 6 10$^5$ K. The non-thermal
velocities  (40 - 45 km/s) were somewhat smaller but close to
 the sonic ones.
The most striking feature of the present observation, lasting 120 min,
 was the strong microflaring  and  the appearance of  short lived transients.
 One of
these   (BP-B in Fig.2) showed clear mass upflows with velocities
20 km/s.   No flows were detected elsewhere in the observed region.

Inside the 8 x 8 arcsec area of the BP-A no clear structure could be
resolved within the  2 arcsec spatial resolution. All spatial
pixels varied  coherently with strong correlation between individual
pixels (Fig. 4).
 It is   possible that the BP-A
consisted of a dense system of small loops  triggering each others,
so that the whole BP looked like a single variable object.
A much better spatial resolution (like in {\it TRACE}, Brown et al. 2001)
is needed to resolve the issue and to find the BP building blocks.
 
Comparing  the sun and solar type active stars, Vilhu (1987, 1994)
suggested
a correlation between the CIV 1550 line intensity (10$^5$ K) and 
the magnetic filling
factor. Lines formed at 10$^5$ K
arise close to the loop foot points and are good indicators of their
surface area. 
In the quiet solar network the magnetic filling factor is at one per cent level
(one per cent  of the surface covered by equipartition fields) while
in the most active (and consequently rapidly rotating) stars the filling
factor approaches 100 per cent  
when the CIV line surface flux  is  100 times
that of the quiet sun. 
This suggests that the filling factor of BP-A was around 10 per cent 
(see Fig. 5).
In this picture  BP-A was
 similar to the average surface of a moderately active star
(like solar type stars in the Hyades cluster, rotating with
periods around 10 days or less).  Further, some bright points 
(like the present one) may well be
essentially like over dense quiet sun regions. However, much better
spatial resolution will be needed to make any definite conclusion.

We also investigated the underlying magnetic field structure of the region
during our observation (from the SOI-archive).
 The MDI-instrument onboard {\it SOHO} observed the
region  between UT 15:16 - 15:36 and 16:45 - 17:05.
In both MDI-observations (around the BP-A of Fig.2)  two  
bright positive flux regions  
(10  and 20 arcsec apart) and two negative flux fragments 
(30 arcsec apart) are present and look rather persistent.
 According to Brown et al. (2001) one third of the
bright points lie over emerging regions of magnetic flux while the remaining
two thirds lie above cancelling magnetic features. It seems that the bright point
studied in the present paper belongs to this latter category.

\begin{acknowledgements}
This work was performed with support from the Academy of Finland (OV and JH).
 We are very grateful to 
Matts Carlsson  for providing 
software and useful information for  basic data reductions.
 We thank the referee for valuable criticism.
\end{acknowledgements}

%
   \begin{figure*}
\resizebox{12cm}{!}{\includegraphics{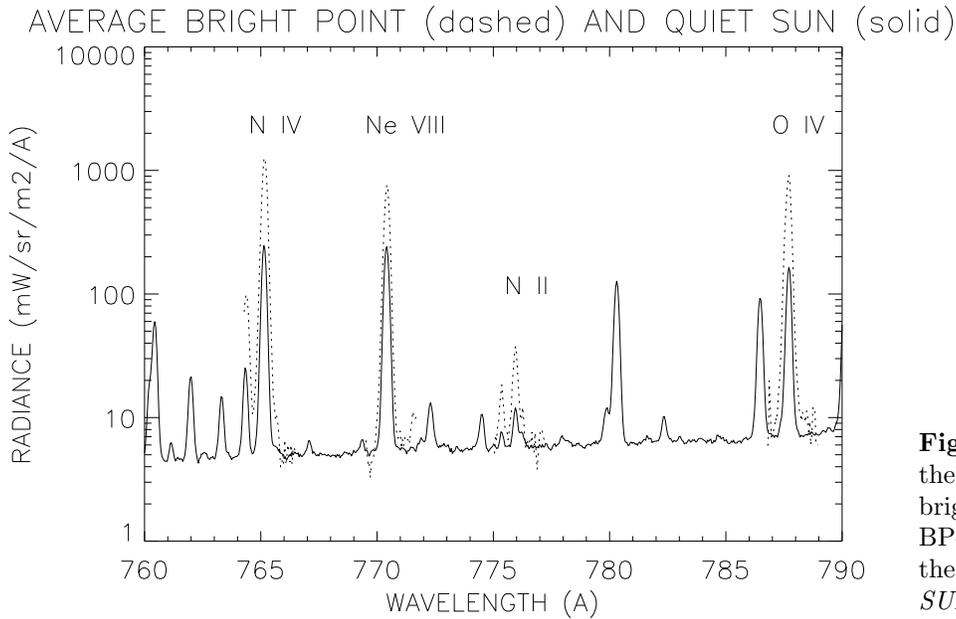}}
\hfill      
\parbox[b]{55mm}{\caption{ The spectral range with the lines used marked. The
average bright point spectra (dashed lines, BP-A in Fig. 2) 
are overplotted with the quiet 
sun spectrum from the {\it SUMER}-atlas (solid line, available via
http://www.mpae.gwdg.de). }}
\label{FigSpec}
\end{figure*}
   \begin{figure*}
\resizebox{12cm}{!}{\includegraphics{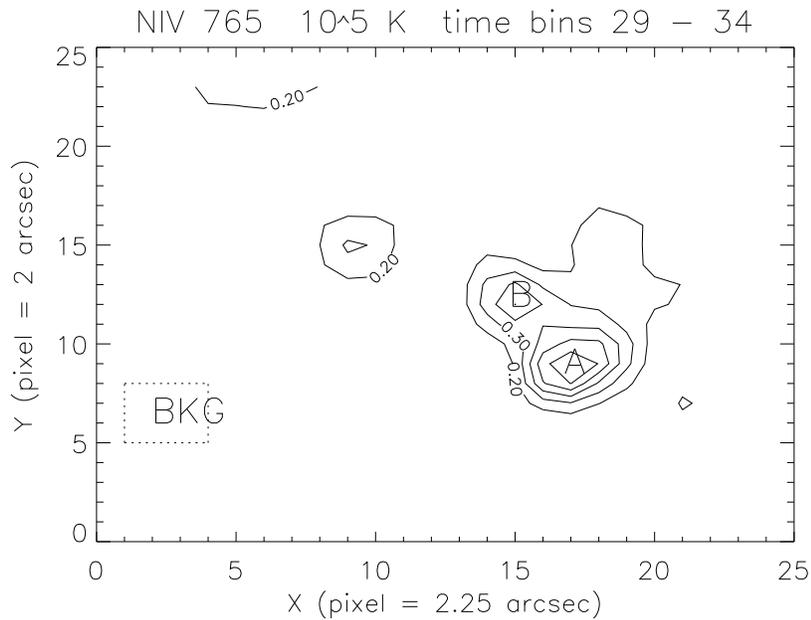}}
\hfill      
\parbox[b]{55mm}{\caption{  NIV 765 contour-map  (W/Sr/m$^2$) 
 of the region observed. The map is an average of six time intervals
(frames 29 - 34 in Fig.3). 
The bright points A and B discussed in the text are marked.
'BKG' refers to the background region used in Figures 4 and 5. The 
X-coordinate (in pixels, one pixel = 2.25 arcsec)
 points to West and Y (in pixels, one pixel = 2 arcsec) to 
North. The map center is at 91 arcsec to West and 15 arcsec to South from
the solar  center in the beginning of observations.
 The region was observed in October 8, 1996,
between UT 15:20 - 17:20 (two hours). }}
\label{FigContour}
\end{figure*}
   \begin{figure*}
\resizebox{12cm}{!}{\includegraphics{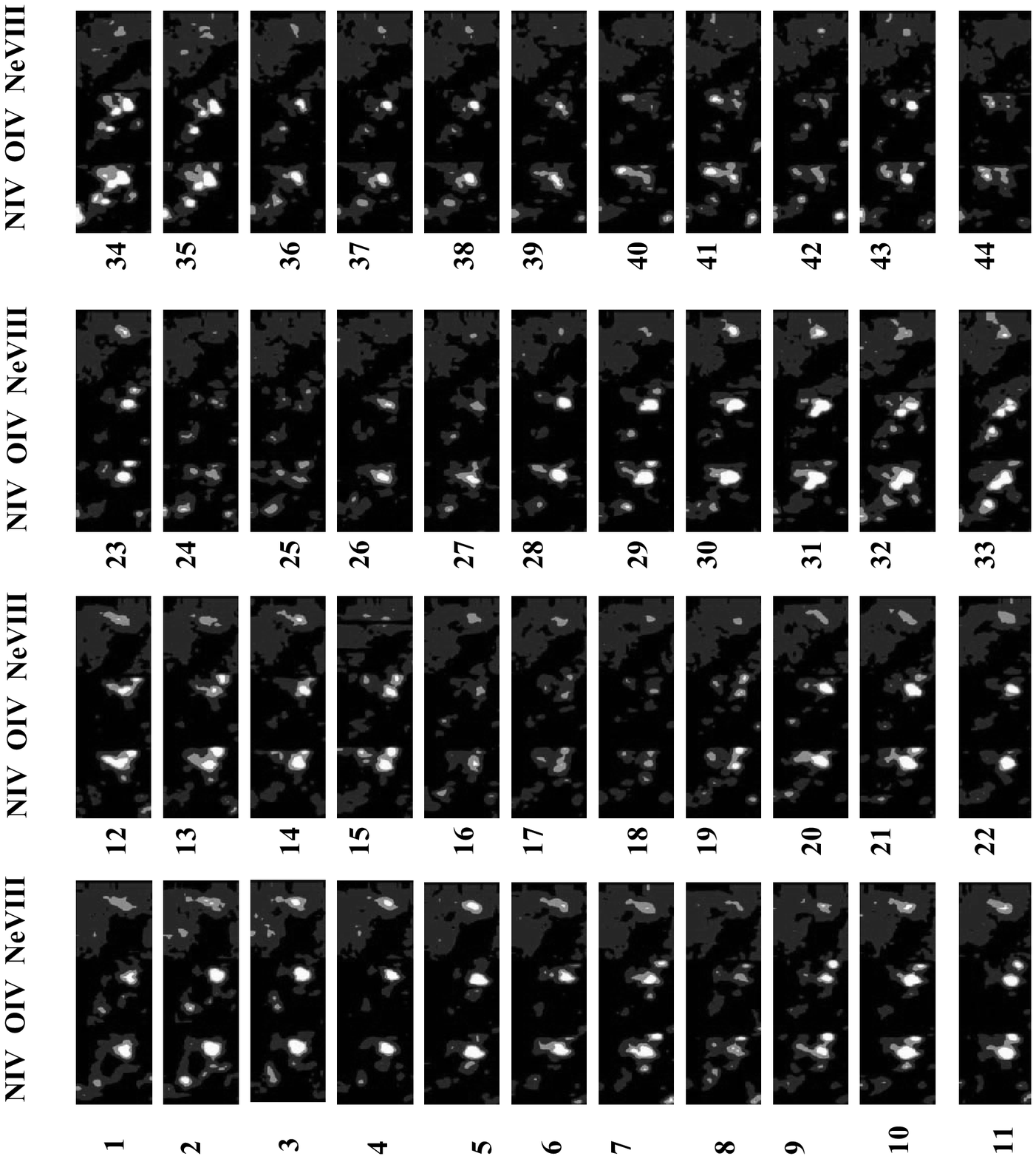}}
\hfill      
\parbox[b]{55mm}{\caption{  The collection of line intensity maps over
the 50 x 50 arcsec region observed.
The  maps are shown for three lines formed at different
temperatures: NIV (10$^5$ K, OIV (2 10$^5$ K) and NeVIII (6 10$^5$ K). 
Every observed frame is shown and labeled. The time difference between
contiguous images is 2.8 minutes. 
 BP-A (see Fig.2) is
dominating during the whole observation while the nearby BP-B 
(see Figs.2 and 6)
is a short lived transient appearing in frames numbers 29 - 34.  }}
\label{FigSequence}
\end{figure*}
   \begin{figure*}
\resizebox{12cm}{!}{\includegraphics{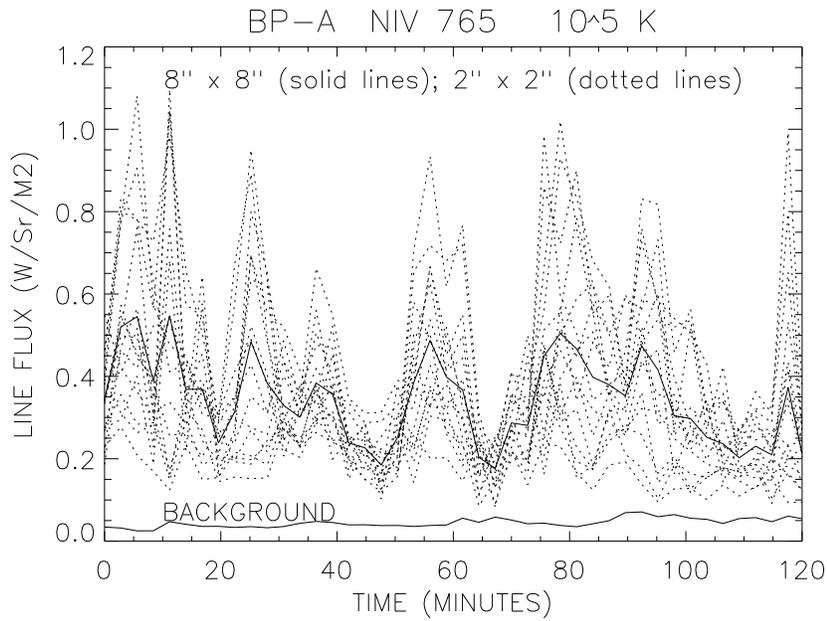}}


\hfill      
\parbox[b]{55mm}{\caption{ The light curve of NIV 765 total intensity 
of the BP-A (see Fig.2)
using 16 spatial pixels (2 x 2 arcsec each) and combining them into a single
(average over 8 x 8 arcsec) light curve (the solid line). 
The light curves of all 16 individual  spatial pixels are shown separately 
by dotted curves.
Note the clear coherence of separate pixels.
The background was computed from the 
8 x 8 arcsec region shown in Fig. 2 as 'BKG'. }}
\label{lightcurveA}
\end{figure*}
%

   \begin{figure*}
\resizebox{12cm}{!}{\includegraphics{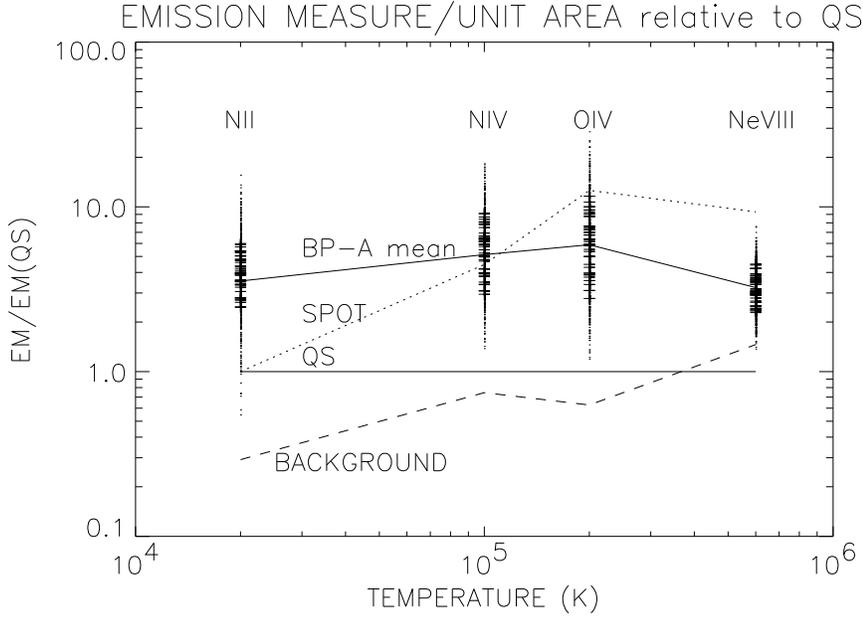}}
\hfill      
\parbox[b]{55mm}{\caption{ Emission measures 
in the BP-A (see Fig.2) vs temperature, in units of 
the quiet sun value (from {\it SUMER} ATLAS, available via 
http://www.mpae.gwdg.de).  
 These relative emission measures were computed simply as ratios of
total line fluxes. They are  proportional to 
n$_e$$^2$ $\times$ V where n$_e$ is the
electron density and V the emitting volume  at the formation
temperature of the line in question.
  The short horizontal ticks give 8 x 8
arcsec averages for separate  44 time bins, while the dots (forming vertical
lines) show all individual 2 x 2 arcsec pixels  for all 44 time bins.
The dashed line marked 'BACKGROUND' was computed from the box shown in Fig. 2,
while the dotted line shows the spot-spectrum from the {\it SUMER} ATLAS.
It is suggested that EM/EM(QS) at 10$^5$ K is briefly equal to the magnetic
flux tube filling factor (in per cent).
                               } }
\label{EM}
\end{figure*}
%

   \begin{figure*}
\resizebox{12cm}{!}{\includegraphics{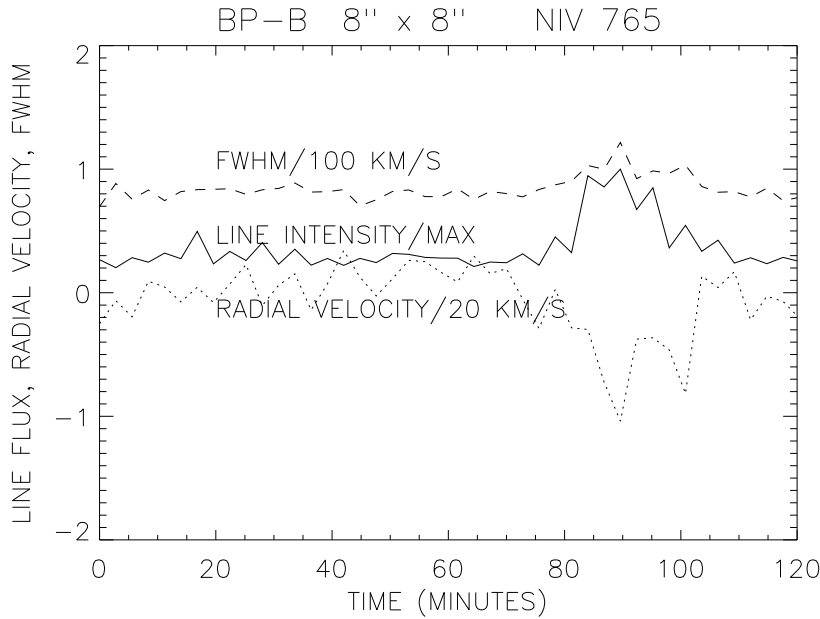}}
\hfill      
\parbox[b]{55mm}{\caption{Average (over 8 x 8 arcsec) light curves
of the short lived 
transient B  (see Fig.2; appearing in frames numbers 29 - 34 of Fig.3)
 for the NIV-line total intensity,
radial velocity (negative means blue-shifted upward motion) and
line width (FWHM). 
The average FWHM is the same as that of the major
bright point A (see Table 1)
and indicates  nonthermal broadening of 40 km/s.
 This dynamic event may be associated
magnetically with the neighbouring BP-A  showing mass-upflows  20 km/s.
The errors of radial velocity and FWHM are around 5 km/s.
                             } }
\label{B}
\end{figure*}

\end{document}